PATRICIA PASKOV, MICHAEL BYUN, KEVIN WEI & TOBY WEBSTER

# Preliminary suggestions for rigorous GPAI model evaluations

April 2025

## I. Introduction

Researchers are increasingly calling attention to shortcomings in general-purpose AI (GPAI) evaluation[1] methodologies[2] that can affect the robustness and reliability of evaluation results.[3] Current evaluations have been criticised for lack of statistical rigour, failure to measure real-world risk, insufficient public documentation, poorly designed constructs and instruments, and other shortcomings.[4] The field of GPAI evaluations would thus benefit from clearly set out best practices guiding the design, implementation, execution and documentation of evaluations. Adjacent fields offer established standards from which to draw.

As our team at RAND conducts evaluations of GPAI models,[5] we are actively considering and developing best practices for their rigorous implementation. We share our considerations in this document, which presents a preliminary compilation of suggestions that may improve the internal validity, external validity and reproducibility of GPAI evaluations. We put forth suggestions for two evaluation types – human uplift studies[6] and benchmark evaluations[7] – as well as cross-cutting suggestions that may apply to many different evaluation types. The suggestions have been collated using a literature review, the methodology of which is

> The science of GPAI evaluations is nascent and evolving, and consensus is lacking on how to robustly measure model capabilities and impacts amidst rapid technological advancement.

detailed in Appendix A, drawing from best practices in machine learning, statistics, psychology, economics, biology and other fields recognised to hold important lessons for GPAI evaluation.[8]

The science of GPAI evaluations is nascent and evolving,[9] and consensus is lacking on how to robustly measure model capabilities and impacts amidst rapid technological advancement. With this in mind, we offer these suggestions as preliminary considerations within an ongoing conversation. The intended audience of this document is technical in nature and includes providers of general-purpose AI models presenting systemic risk (GPAISR),[10] for whom the EU AI Act[11] lays out specific evaluation requirements; third-party evaluators; and academic researchers developing or conducting model evaluations.[12] Rather than covering each suggestion in depth, we provide a selection of relevant research for readers to explore further. This format enables our target audience to quickly navigate evaluation approaches and identify entry points for deeper exploration.

The remainder of this document is structured as follows: Section II introduces the elements of high scientific and technical rigour that our suggestions seek to promote, presents the evaluation types in scope, and outlines the four evaluation life cycle stages by which we organise suggestions. Section III presents our suggestions for rigorous GPAI model evaluations. Section IV concludes the report.

## II. Motivation, framework and definitions

In this section, we provide policy-relevant motivation for our suggestions and define the elements of high scientific rigour that they seek to promote (Section II.a). We then present the evaluation types in scope (Section II.b), and outline the four evaluation life cycle stages by which we organise suggestions (Section II.c).

### II.a Elements of high scientific and technical rigour

The suggestions in this piece aim to promote the internal validity, external validity and reproducibility of GPAI evaluations, motivated in part by the European Union (EU) AI Act mandate that providers of general-purpose AI models presenting systemic risk (GPAISR) must conduct evaluations on those GPAISRs.[13] The March 2025 draft of the Code of Practice, which details the AI Act rules, further specifies that GPAISR providers shall ensure that relevant model evaluations are of high scientific and technical rigour:[14] that is, that they should demonstrate high levels of internal and external validity and appropriate levels of reproducibility.[15]

GPAI evaluations fulfilling these criteria may provide decision makers and policymakers with more accurate information about GPAI risks and, in turn, increase the window of opportunity for effective mitigations and preparedness. As such, in seeking to promote the rigour of GPAI evaluations, the suggestions in this document also aim to strengthen downstream societal resilience more broadly. Table 1 defines the elements of high scientific and technical rigour that our suggestions seek to promote.

## II.b Evaluation types in scope

We put forth suggestions for two evaluation types – human uplift studies and benchmark evaluations – as well as cross-cutting suggestions that may apply to many different evaluation types.[16] Human uplift studies measure the extent to which access to and/or use of a GPAI model, relative to status quo tools (e.g. internet search), impacts human performance on a task. Benchmark evaluations consist of a standardised set of items on which GPAI model responses or behaviour are graded. We provide extended definitions for these evaluation types in Sections III.b and III.c, respectively.

## II.c Evaluation life cycle stages in scope

For each evaluation type, we organise suggestions according to four stages in the evaluation life cycle: design, implementation, execution and documentation. We focus on a subset of the life cycle model outlined by Reuel et al. (2024a), originally designed for benchmark evaluations.[17] Here we adapt Reuel et al. (2024a)'s definitions to a broader set of GPAI evaluations, including human uplift studies, and add an execution stage to account for additional methodological considerations presented by non-benchmark evaluations. Table 2 defines the four life cycle stages by which we organise suggestions.

**Table 1: Elements of high technical and scientific rigour**

| | |
|---|---|
| Internal validity | The extent to which evaluation results capture the truth in the evaluation setting – namely, the cause-effect relationship between the subject of interest and the outcome metrics – and are not due to methodological shortcomings.[18] |
| External validity | The extent to which evaluation results can be generalised to model behavior in contexts outside of the evaluation environment.[19] |
| Reproducibility | The extent to which consistent evaluation results can be obtained using the same input data, computational methods, code and evaluation conditions, allowing for other researchers and engineers to validate, reproduce or improve on evaluation results.[20] |

**Table 2: GPAI evaluation life cycle stages, adapted from Reuel et al. (2024a)**

| Design | Implementation | Execution | Documentation |
|---|---|---|---|
| The definition of an evaluation's purpose, scope and structure. At the design stage, evaluators define the key aspects of the AI model(s) that the evaluation seeks to assess and select the appropriate tasks and evaluation metrics to be used in the evaluation. | The selection and/or construction of evaluation tools, including datasets (existing or custom-built), evaluation scripts, APIs and prompting techniques. For research involving human participants, implementation also includes recruiting participants, designing user interfaces and establishing policies to maintain experimental integrity. This stage concludes with the initial testing and validation of evaluation tools. | The process of running the evaluation and collecting results such as model outputs, baselines and human participant data. At the execution stage, evaluators use the tools built or selected during implementation to run the evaluation according to the specified research design. This stage includes collecting and grading model outputs, calculating metrics and analysing evaluation results. | The recording and sharing of evaluation results, metrics, methodological details, experimental materials and datasets to relevant audiences,[21] which may include internal stakeholders, external validators, research partners, government agencies[22] or the general public.[23] |

## III. Preliminary suggestions for rigorous GPAI model evaluations

In this section, we provide suggestions for rigorous GPAI evaluations. We include suggestions that apply across many evaluation types (Section III.a), as well as specific guidance for human uplift studies (Section III.b) and benchmark evaluations (Section III.c).[24] Our suggestions draw from a literature review of 64 articles about AI evaluation methodology, as well as literature on measurement science in fields recognised to hold important lessons for GPAI evaluations.[25] Appendix A provides further details on the review methodology. As a result of this process, some suggestions are highly specific to GPAI evaluations, while others are more broadly applicable scientific standards that are yet to be widely adopted in GPAI evaluations.

Within each sub-section below, we organise suggestions by evaluation life cycle stage. We tag each suggestion with the element(s) of high scientific and technical rigour it may promote, as defined in Table 3. Appendix B details the tagging criteria for each element. Three authors independently categorised each suggestion using the definitions in Appendix B. They discussed any disagreements to reach consensus, with an external expert validating and resolving any remaining non-unanimous cases. Two authors then collaboratively classified suggestions into the evaluation lifecycle stages as defined in Section II.c.

**Table 3: Tags for high scientific and technical rigour**

| | |
|---|---|
| **I** | Suggestion may promote internal validity |
| **E** | Suggestion may promote external validity |
| **R** | Suggestion may promote reproducibility |

### III.a Cross-cutting suggestions for rigorous GPAI model evaluations

In this section, we provide cross-cutting suggestions for rigorous model evaluations. Across many evaluation types, including human uplift studies and benchmark evaluations, our suggestions include:

## Design

| | | | |
|---|---|---|---|
| **Specifying the research question**, including the definition of concepts, capabilities or characteristics to be measured and the context to which the evaluation results aim to generalise.[26] | I | E | |
| **Drawing from expertise** specific to the evaluated domain when conducting risk identification,[27] metrology[28] and evaluation design. | I | E | |
| **Validating evaluation items and grading criteria with domain experts** to ensure that items are topical, possible to complete with the provided information and tools, and at the desired level of difficulty; and that grading criteria are appropriate.[29] | I | E | |
| **Describing and/or validating the extent to which an evaluation method does and does not measure the desired concept**, capability or characteristic.[30] | I | E | |
| **Defining the evaluation environment**, affordances to be provided to the evaluated models, and, where relevant, sources of data used for the evaluation. | I | E | R |
| **Defining continuous or subdivisible performance metrics where possible**, rather than relying on binary success/fail outcomes.[31] | I | | R |
| **Conducting power analyses** to determine appropriate sample sizes according to minimum detectable effect sizes, the underlying distribution of the outcome metric, and Type I and Type II error rates.[32] In the case of novel evaluations and outcome metrics, knowledge of the underlying distribution of the outcome metric and the likely expected effect size – and subsequently, of what would constitute a meaningful effect – may not exist. In these cases, evaluators may consider conducting power analyses for multiple scenarios (e.g. small, medium and large effect sizes) or using the most relevant previous research to estimate potential effect sizes for the purposes of power analysis.<br>• For benchmark evaluations, power analyses may inform the appropriate number of runs and/or questions in the benchmark dataset.<br>• For human uplift studies, power analyses may inform treatment and control group sizes.<br>• For evaluations to which power analyses do not apply, suggestions include justifying the absence of power analyses. | I | E | |
| **Running the evaluation with a sufficient sample size,** as indicated by power analyses. | I | E | |
| **Drawing up a pre-analysis plan** to increase transparency and rigour, with pre-registration as the gold standard.[33] Pre-analysis plans may include details on sample size, randomisation strategy where relevant, data cleaning, data analysis, metric construction, statistical testing, and reference points to which metrics will be compared (e.g. pre-defined capability thresholds). | I | | R |
| **Estimating the statistical uncertainty** of results, where relevant, using methods appropriate for the size and structure of the dataset, including confidence intervals, standard errors, hypothesis testing[34] and Bayesian statistics.[35] | I | E | |
| *For research including human participants:* | | | |
| **Ensuring ethical human subjects research practices**, including but not limited to informed consent; privacy-preserving data collection and data storage protocols; and, more broadly, Research Ethics Board or Institutional Review Board approval. | I | | |
| **Specifying a population of interest** through axes relevant to the research question, including but not limited to domain expertise, age, education, ethnicity, gender, geographic location, language, nationality and socioeconomic status.[36] | | E | R |
| **Developing a sampling and recruitment strategy** with appropriate inclusion/exclusion criteria for selecting human participants representative of the population of interest.[37] | | E | R |
| **Complementing quantitative data with qualitative data collection** (e.g. interviews, surveys and chat logs) to more comprehensively understand the experiences of participants, the nature of their interactions with GPAI models, and the mechanisms through which evaluation results occur.[38] | I | | |
| **Developing a strategy for incentivising human participants to perform tasks as desired** in an ethical manner that minimises bias and maintains, to the extent possible, a sample representative of the population of interest.[39] | I | E | |

## Implementation

| | | | |
|---|---|---|---|
| **Specifying and/or designing performance elicitation techniques and tools**, including enhancements in tooling, prompting, scaffolding, solution choice and fine-tuning,[40] to capture evaluation results that respond to the specified research question. When seeking to measure capability ceilings that may appear in deployment settings, for example, evaluators may use these techniques and tools to attempt to elicit maximum capabilities. As a gold standard, evaluators may ensure that the individuals performing elicitation are blinded to the design, scoring criteria and answer key of the evaluation itself. | I | E | R |
| *For research including human participants:* | | | |
| **Specifying the protocols by which participants can access and leverage GPAI tools**, with attention to the amount of time and compute available. | I | E | R |

## Execution

| | | | |
|---|---|---|---|
| **Conducting sensitivity analyses**[41] **and ablation experiments**[42] to measure the extent to which factors like prompt wording, scaffolding and choice of elicitation techniques affect results. | I | E | |
| **Monitoring and checking chain-of-thought traces** for reasoning that may confound results, including model awareness of being under evaluation[43] or 'sandbagging' (i.e. a model downplaying its capabilities).[44] | I | E | |
| **Controlling for or constraining variables such as compute budget**[45] **and time** when comparing results across models and evaluation runs. | I | E | R |
| *For research including human participants:* | | | |
| **Monitoring human participants' compliance with experimental conditions**, including by running attention checks and/or measuring adherence to policies specifying acceptable GPAI model use. | I | | R |

## Documentation

| | | | |
|---|---|---|---|
| **Documenting whether and how each suggestion in Design, Implementation, and Execution was performed**. | | | R |
| **Documenting data sources and collection, processing and annotation methods**, including how evaluation tasks were collected or created; filtered and edited; and how gold standard answers were determined.[46] | | | R |
| **Documenting how an evaluation result should or should not be interpreted**.[47] | | | R |
| **Documenting model parameters used in evaluations**,[48] including model version(s), model quantisation(s) and hyperparameters such as temperature and limits on length of outputs.[49] | | | R |
| **Securely releasing execution details**, including evaluation and grading methodology, computational environment (e.g. hardware architecture, cloud hosting decisions, operating systems and library dependencies), prompting and elicitation methods.[50] | | | R |
| **Securely releasing evaluation code**,[51] including that for randomisation, data cleaning, statistical testing and analysis. | | | R |
| **Documenting the level and duration of access** to GPAI models provided for the purpose of the evaluation.[52] | | | R |
| **Documenting the calendar time, compute budget, personnel count and hours-per-person** spent designing, implementing and executing the evaluation. | | | R |
| **Documenting the distribution of results** across participants, models, runs and/or other variables of interest, including with instance-level outputs.[53] | | | R |
| *For research including human participants:* | | | |
| **Documenting anonymised participant recruitment and participation details**, including relevant demographics (e.g. domain expertise, previous experience with GPAI systems, age, education, geography, nationality, ethnicity, gender, etc.), breakdowns by treatment, attrition, and notes on the degree of representativeness of the population of interest. | | | R |
| **Documenting instructions and/or training** provided to participants, study monitors, data annotators and/or graders. | | | R |

## III.b Suggestions for rigorous human uplift studies

In this section, we provide suggestions for rigorous human uplift studies. Human uplift studies measure the extent to which access to and/or use of a GPAI model, relative to status quo tools (e.g. internet search), impacts human performance on a task. Human uplift studies often employ randomised controlled trial design to form a grounded assessment of the causal impact of an GPAI system on human performance.[54] Previous examples of human uplift studies include those by RAND,[55] OpenAI[56] and Anthropic,[57] all of which exhibit methodological limitations despite their contributions.[58]

Our suggestions for rigorous human uplift studies include:

### Design

| | | | |
|---|---|---|---|
| **Conducting stratified or clustered randomisation** for treatment and control groups, where appropriate,[59] on demographics critical to the research question of interest, including but not limited to domain expertise and prior experience with GPAI systems; and ensuring randomisation is reproducible (e.g. by setting a random seed). | I | | R |
| **Designing treatment and control conditions to mitigate confounding variables** by ensuring, to the extent possible, that experimental conditions between groups are identical except for access to the GPAI model(s) and tools being studied.[60] | I | E | R |
| **Carefully designing user interfaces (UI)** that appropriately respond to the research question and remain constant across treatment and control groups. If seeking to evaluate the impact of state-of-the-art tools, for example, evaluators may consider providing state-of-the-art UI to all participants. Low-quality UI across groups may lead to frictions that falsely present as low performance; similarly, if treatment groups receive tooling with a superior UI than that provided to control groups, this disparity may lead to lower control group performance and a subsequent overestimation of human uplift attributable to the GPAI model itself. | I | E | R |
| **Carrying out the cross-cutting suggestions included in Section III.a: Design.** | I | E | R |

## Implementation

| | | | |
|---|---|---|---|
| **Piloting/validating survey instruments and experimental protocols** (e.g. infrastructure, tool use and data quality monitoring mechanisms) with out-of-sample participants who resemble the sample.[61] | I | E | |
| **Clearly defining the participants' access to tooling** with a careful and balanced consideration of both the research question and the degree to which diffusion of AI may be captured in control conditions[62] and may thus lead to an underestimation of human uplift.[63] | I | E | R |
| **Designing frameworks and collecting data to monitor and control for non-compliance** (e.g. of control or treatment group tool use) and contamination; and to more granularly measure and analyse mechanisms by which outcomes occur.[64] | I | E | R |
| **Carrying out the cross-cutting suggestions included in Section III.a: Implementation.** | I | E | R |

## Execution

| | | | |
|---|---|---|---|
| **Running high frequency data quality checks**, including attention checks,[65] to monitor and optimise data quality during the course of the evaluation.[66] | I | | |
| **Ensuring that all graders are blinded** to the treatment category when reviewing outputs;[67] and furthermore assessing the quality of the blinding.[68] | I | | |
| **Carrying out the cross-cutting suggestions included in Section III.a: Execution.** | I | E | R |

## Documentation

| | | |
|---|---|---|
| **Documenting whether and how each suggestion included in Design, Implementation, and Execution was performed**. | | R |
| **Carrying out the cross-cutting suggestions included in Section III.a: Documentation**. | | R |

## III.c Suggestions for rigorous benchmark evaluations

In this section, we provide suggestions for rigorous benchmark evaluations. Benchmark evaluations consist of a standardised set of items on which model responses or behaviour are graded. Benchmarks allow comparison – at scale and over time – of model performance to that of other models or humans. Increasingly, researchers and policymakers interpret benchmark results with reference to specific capability thresholds, such that model performance approaching or crossing a threshold may trigger policy decisions, as outlined in various Frontier Safety Frameworks.[69] Benchmark evaluations are designed to be easily repeatable and can provide early evidence of capabilities and risks upon which further in-depth methods such as human uplift studies can build. Current benchmark evaluations vary along a number of dimensions:

- Output format:
    - **Multiple-choice question answering (MCQA) benchmark evaluations:** contain items that prompt a model with a question or task and a small set of candidate answers,[70] evaluating the quality of a model's selection (single or multiple) from the answer set. Examples include Massive Multitask Language Understanding (MMLU)[71] and a Graduate-Level Google-Proof Q&A (GPQA).[72]
    - **Open-ended benchmark evaluations**[73]: contain items that prompt a model with an open-ended question or task, evaluating the quality of a model's *generated output,* whether in a question-answer (QA) format or a non-QA, short-form, task format. Examples include MATH[74] for QA format, and HumanEval[75] for short-form, non-QA task format.[76]
    - **Long-form task**[77] **benchmark evaluations**: contain items that prompt a model to perform a more complex task. Compared to MCQA and open-ended benchmark evaluations, long-form task-based benchmark evaluations usually have a longer horizon length: that is, they require more time, effort, compute or other resources (e.g. internet search) for both humans and models to complete,[78] often consisting of multiple steps in pursuit of a goal. The final output may take the form of a simple one-word answer, a complex artifact (e.g. a piece of code) or dynamic performance in a game-like environment. Examples include ARA evaluations,[79] PaperBench,[80] SWE-bench,[81] MLAgentBench,[82] MLE-bench[83] and Capture the Flag (CTF) challenges.[84]
- Grading strategy: grading may be automated (e.g. exact or fuzzy matching, unit tests for code),[85]

expert or non-expert human-graded (e.g. helpfulness grading, summarisation quality),[86] model-graded,[87] or a combination of these strategies. GPAI evaluation best practices may vary in importance depending on grading strategy: for example, exact matching may warrant more validation to mitigate external validity issues,[88] while human grading requires best practices related to human participants.

- Size: benchmark evaluations range from small and bespoke (e.g. the seven engineering tasks in RE-Bench)[89] to very large (e.g. the 14 million image-caption pairs in ImageNet).[90] Some suggestions, such as the choice of statistical method (e.g. for uncertainty estimation), vary depending on the size of the benchmark.

Benchmark evaluations have been extensively studied in the GPAI evaluations literature. Reuel et al. (2024a), in particular, propose comprehensive recommendations for best practices from which we draw.

Our suggestions for rigorous benchmark evaluations include:

### Design

| | | | |
|---|---|---|---|
| **Carrying out the cross-cutting suggestions included in Section III.a: Design.** | I | E | R |

### Implementation

| | | | |
|---|---|---|---|
| **Testing if the benchmark evaluation can be solved by shortcuts**, for instance by removing necessary information.[91] | I | E | |
| **Carrying out the cross-cutting suggestions included in Section III.a: Implementation.** | I | E | R |

### Execution

| | | | |
|---|---|---|---|
| **Using a validation or test set** that is not released publicly.[92] | I | E | |
| **Using methods to measure the extent to which the model has been trained** on data from the benchmark.[93] | I | E | |
| **Establishing robust and well-documented human baselines**[94] where relevant.[95] | | E | |
| **Carrying out the cross-cutting suggestions included in Section III.a: Execution.** | I | E | R |

**Documentation**

| | | |
|---|---|---|
| **Documenting whether and how each suggestion in Design, Implementation, and Execution was performed**. | | R |
| **Documenting the metric of interest** (e.g. accuracy, pass@k, average quality score) and the methodology by which it is measured.[96] | | R |
| **Documenting reference scores**, including, where available, floors (i.e. minimum performance), ceilings (i.e. maximum performance), human baselines and/or random performance levels for the chosen metric(s) to further assist users in understanding a model's score.[97] | E | R |
| **Using a globally unique identifier (GUID) or canary string**, or encrypting public evaluation code and data.[98] | | R |
| **Securely releasing the evaluation** via software packages and code bases that researchers can easily use without training,[99] including a requirements file, script to replicate results, and code documentation.[100] | | R |
| **Documenting test environment design and prompt design** process.[101] | | R |
| **Documenting representative samples** of benchmark items and model responses.[102] For risk-relevant evaluations, summary statistics may be provided for underlying items and/or samples may be provided over proxy tasks. | | R |
| **Carrying out the cross-cutting suggestions included in Section III.a: Documentation.** | | R |

# IV. Conclusion

This document outlines suggestions that may promote the internal validity, external validity and reproducibility of GPAI evaluations. These suggestions, motivated by our team's work designing and running evaluations of GPAI models, draw from the fields of machine learning, statistics, psychology, economics, biology and other fields recognised to have important lessons for GPAI evaluation. We focus in particular on methodological concerns during the design, implementation, execution and documentation of human uplift studies and benchmark evaluations, as well as on cross-cutting suggestions that may apply across evaluation types. Although these suggestions are preliminary and high-level, we hope they may help shape discussions about building and running GPAI evaluations, particularly as providers of GPAISR seek to integrate the EU AI Act Code of Practice call for rigorous evaluations into their work. Future research may explore suggestions across an expanded scope of evaluation types, prioritise suggestions through cost-benefit analyses, and validate and refine the role of these suggestions within the GPAI evaluation ecosystem.

## Notes

[1] Defined by the EU AI Act as "an AI model, including where such an AI model is trained with a large amount of data using self-supervision at scale, that displays significant generality and is capable of competently performing a wide range of distinct tasks regardless of the way the model is placed on the market and that can be integrated into a variety of downstream systems or applications" (European Parliament and Council of the European Union, 2024). The EU AI Act excludes from its definition of GPAI models "AI models that are used for research, development or prototyping activities before they are placed on the market." "GPAI model" is often used synonymously with the term "foundation model."

[2] Defined as the empirical assessment of the 'components, capabilities, behavior, and impact' of an AI model (Weidinger et al. 2024). In this piece, we use GPAI evaluation as an umbrella term for alignment, capability, propensity and safety evaluations, all of which may benefit from the suggestions outlined in Section III.

[3] E.g. Biderman et al. (2024); Eriksson et al. (2025); Liao et al. (2021b); Rauh et al. (2024); Subramonian et al. (2023); Wallach et al. (2024); Weidinger et al. (2025).

[4] Cowley et al. (2022); Eriksson et al. (2025); Liao et al. (2021b); McIntosh et al. (2024); Reuel et al. (2024a); Tedeschi et al. (2023).

[5] E.g. Mouton et al. (2024); Dev et al. (2025); Persaud et al. (2025).

[6] Human uplift studies measure the extent to which access to and/or use of a GPAI model, relative to status quo tools (e.g. internet search), impacts human performance on a task. We provide an extended definition of human uplift studies in Section III.b.

[7] Benchmark evaluations consist of a standardised set of items on which GPAI model responses or behaviour are graded. We provide an extended definition of benchmark evaluations in Section III.c.

[8] Reuel et al. (2024a); Wallach et al. (2024); Weidinger et al. (2025).

[9] Apollo Research (2024); Bengio et al. (2025); Kapoor et al. (2024); Reuel et al. (2024b); Weidinger et al. (2025).

[10] A GPAI model is considered by the EU AI Act as presenting systemic risk if it either a) "has high impact capabilities evaluated on the basis of appropriate technical tools and methodologies, including indicators and benchmarks" or b) "based on a decision of the Commission, ex officio or following a qualified alert from the scientific panel, it has capabilities or an impact equivalent to those set out in point (a)" (European Parliament and Council of the European Union (2024)).

[11] The EU AI Act is the first-ever legal framework on AI, entering into force on 1 August 2024. Its governance rules and obligations for general-purpose AI models become applicable on 2 August 2025. The AI Act seeks to "improve the functioning of the internal market and promote the uptake of human-centric and trustworthy artificial intelligence (AI), while ensuring a high level of protection of health, safety, fundamental rights enshrined in the Charter, including democracy, the rule of law and environmental protection, against the harmful effects of AI systems in the Union and supporting innovation" (European Parliament and Council of the European Union (2024)).

[12] The document includes some technical language assumed to be familiar to the intended audience.

[13] See Article 55 in European Parliament and Council of the European Union (2024).

[14] European Commission (2025, Measure II.4.5).

[15] See the Safety and Security Section Glossary in European Commission (2025).

[16] Future work may include content on other evaluation categories such as red-teaming, adversarial testing, model organisms, simulations and proxy evaluations, all of which are listed in the EU AI Act Code of Practice as potential methods for state-of-the-art model evaluations (European Commission (2025)).

17 Reuel et al. (2024a) define five stages of the benchmark life cycle: design, implementation, documentation, maintenance and retirement. In this document, we primarily focus on life cycle stages that most directly inform the methodology and immediate interpretation of evaluations.

18 Patino & Ferreira (2018); European Commission (2025).

19 European Commission (2025); Paskov et al. (2024); Liao et al. (2021a).

20 European Commission (2025); National Academies of Sciences, Engineering, and Medicine (2019).

21 Not all documentation should or must be made public, for example due to security concerns with information sharing. The degree of security required for individual documentation items may vary with factors such as risk domain, level of detail and the nature of the evaluation. Strategies for secure documentation and dissemination may include, for example, tiered-access frameworks (e.g. as proposed in Frontier Model Forum (2025a)) and secure information-sharing agreements (e.g. Frontier Model Forum (2025b); US Department of Commerce (2024)).

22 The EU AI Act Code of Practice Measure II.8.3.10, for example, requires non-small-medium-enterprise (SME) signatories to provide documentation of the internal validity, external validity and reproducibility of their GPAISR evaluations.

23 Documentation and dissemination strategies may aim to promote transparency, accountability and open research while appropriately managing and mitigating risks related to, for example, information security, evaluation contamination, potential for misuse and premature disclosure of vulnerabilities.

24 The decision on which evaluations to use under which circumstances is out-of-scope for this document, and relies on a range of factors including the magnitude of risk, available resources, timeline and results of previous evaluations. According to these factors, evaluators may use a structured approach to prioritise between evaluation types, tasks and constructs for evaluation – both with respect to *whether* and *how* to run a particular evaluation type.

25 An increasing literature has recognised that AI evaluations can learn from measurement science in other fields (e.g. Chouldechova et al. (2024); Reuel et al. (2024a); Wallach et al. (2025); Weidinger et al. (2025)).

26 Chouldechova et al. (2024); Reuel et al. (2024a); Wallach et al. (2024); Weidinger et al. (2025).

27 National Institute of Standards and Technology (2024).

28 Chouldechova et al. (2024).

29 Chowdhury et al. (2024); Hardy (2025).

30 That is, breaking down the concept, capability or characteristic into a clear and structured formulation – and mapping it to the evaluation type – in a manner that enables scrutiny and facilitates understanding among those approaching it with different beliefs or definitions. See, for example, Wallach et al. (2024). In this process, evaluators may consider the extent to which extraneous demands within the evaluation – such as stylistic constraints on output format, as highlighted in Hu et al. (2024) – may mask the capability of interest.

31 Gekker et al. (2025).

32 Cohen (2013).

33 Center for Open Science (n.d.); J-PAL (2023).

34 Miller (2024).

35 Bowyer et al. (2025).

36 Wei et al. (2025).

37 Berinsky (2017); Findley et al. (2021); Stantcheva (2023).

38 O'Cathain et al. (2013); Weidinger et al. (2023). Qualitative data can uncover hidden mechanisms, help identify otherwise unobserved confounding variables, and provide insight into potential under- or overestimations of capability.

[39] Moodley & Myer (2003). Structuring compensation as a base payment plus bonus may effectively reduce attrition and incentivise participants to stay on task. Bonuses may be used to compensate for levels of engagement (e.g. whether a participant reaches a certain check-point) as well as outcome success (e.g. how well the participant completes the task). For longer-form tasks, providing incentives for the completion of sub-sessions can help maintain participant attention.

[40] Davidson et al. (2023).

[41] Reuel et al. (2024a).

[42] E.g. Weidinger et al. (2025).

[43] Hobbhahn (2025).

[44] Gasteiger et al. (2025).

[45] E.g. for evaluations using elicitation and/or agent evaluations.

[46] Reuel et al. (2024a).

[47] Reuel et al. (2024a). For example: 'A high HumanEval pass@k score indicates the model can produce functionally correct solutions for discrete, well-defined programming problems, but does not evaluate code maintainability, security, or performance when integrated into larger systems.'

[48] Liao et al. (2021b).

[49] International Network of AI Safety Institutes (2024).

[50] Paskov et al. (2024).

[51] Goodman (2016).

[52] National Institute of Standards and Technology (2024).

[53] Burnell et al. (2023).

[54] AI Safety Institute (2024); Frontier Model Forum (2024).

[55] Mouton et al. (2024).

[56] Patwardhan et al. (2024).

[57] Anthropic (2025).

[58] Peppin et al. (2024).

[59] Athey & Imbens (2016).

[60] Wei et al. (2025).

[61] Bjarkefur et al. (2021).

[62] For instance, while human uplift studies have traditionally granted control participants access to internet search engines like Google, many of these tools now include AI-generated answers. As such, control group use of AI-generated search answers may lead to an underestimate of AI impacts. With these considerations in mind, researchers should carefully define the research question and design the research methodologies accordingly.

[63] Wei et al. (2025).

[64] Bjarkefur et al. (2021). Minimally, researchers can ask participants following the study whether they used AI tools.

[65] Using LLMs to check for participant engagement is a low-cost and high-return method of measuring data quality.

[66] Bjarkefur et al. (2021).

[67] Day & Altman (2000).

[68] Savović et al. (2012).

[69] METR (2025).

[70] If the logits of a model are exposed (that is, if one can record the specific probabilities that the model assigns to different completions of a prompt), a multiple-choice benchmark may also evaluate the length-normalised log probability that a model assigns to several candidate answers.

[71] Hendrycks et al. (2020).

[72] Rein et al. (2023).

[73] Also known as free-response evaluations.

[74] Hendrycks et al. (2021).

[75] Chen et al. (2021).

76 There is no consensus on whether non-QA tasks belong in this category. However, we posit that there is no clear line between open-ended QA and non-QA short-form tasks. For example, consider the QA prompt 'What does the following sentence mean, in English?…' versus the near-identical, 'task-based' prompt 'Please translate the following sentence to English:…'. The most meaningful distinction along this axis between benchmarks is whether they assess the model's evaluative versus generative capabilities. Non-multiple-choice benchmarks assess the latter, regardless of whether they do or don't involve question-answering.

77 Also known as agent, tool use or task-based evaluations, each referring to a distinct but overlapping characteristic. 'Task-based' evaluations seek to contrast with 'question answering' or short-form evaluations; 'agent' describes the subject of an evaluation (e.g. a model with access to tooling); and 'tool use' indicates the affordances provided to an evaluated model. These terms are not always mutually exclusive to each other or other evaluation types: for instance, an agentic model could use tools such as web search to answer questions on an MCQA benchmark. In addition to these reasons, we choose not to use 'task-based' given its murky definition in the context of AI evaluations: 'task' can refer to an individual benchmark item, a cluster of benchmark items, a category of possible model objectives, or specifically non-QA objectives.

78 Kwa et al. (2025).

79 Anthropic (2025).

80 Starace et al. (2025).

81 Chowdhury et al. (2024).

82 Huang et al. (2024).

83 Chan et al. (2025).

84 OpenAI (2024).

85 While automated grading can be simple, as in the case of grading simple yes/no answers, it is often not straightforward. For example, grading code with naive fuzzy string matching methods such as BLEU score may fail to accurately measure the functional correctness of code.

86 Manual human grading can be time and cost intensive, particularly for expert grading, and requires careful methodology to ensure high-quality, representative data.

87 Gu et al. (2024).

88 Chen et al. (2021).

89 Wijk et al. (2024).

90 Deng et al. (2009).

91 Gururangan et al. (2018); Schlangen (2020). For example, testing whether benchmarks seeking to measure multimodal capabilities can be solved on text alone and without reference to images.

92 Partnership on AI (2024). The benefits of public vs. private test sets depend on the intent of the evaluation; for GPAISR, the net benefit of public sets is questionable.

93 Reuel et al. (2024a); Zhang et al. (2024).

94 Defined as reference sets of metrics intended to represent human performance on specific tasks. Human baselines help to contextualise GPAI evaluation results.

95 A checklist for human baseline best practices is provided in Wei et al. (2025).

96 Reuel et al. (2024a).

97 Reuel et al. (2024a).

98 Reuel et al. (2024a).

99 Hendrycks & Woodside (2024).

100 Reuel et al. (2024a).

101 Reuel et al. (2024a).

102 As in, for example, Philip & Hemang (2024) and Phan et al. (2025).

# Appendix A: Literature review methodology

## GPAI evaluations literature review

We performed a three-step literature review of 64 papers on GPAI evaluations:

1. We searched Google Scholar and Advanced Google Search for papers published between 1 January 2020 and 9 December 2024 that fulfilled three criteria. The search returned 85 papers. The criteria were:
   - contains: 'capability evaluation' OR 'alignment evaluation' or 'safety evaluat*' AND
   - contains at least one of the words: taxonomy classification methodology framework review categorisation benchmark assessment evaluation examin* AND
   - contains the exact phrase: 'foundation model' OR 'general purpose AI' OR 'generative AI' OR 'large language model' OR 'LLM' OR 'genAI'.
2. We filtered the results by abstract to retain papers that focused on methodologies and taxonomies of GPAI evaluations. 35 papers remained. Papers were removed if:
   - the source was a personal blog or forum
   - the paper did not address GPAI evaluations
   - the paper focused on the execution of one specific benchmark rather than broader methodological approaches.
3. Through snowball sampling of filtered papers, 29 additional papers were added.

## Interdisciplinary knowledge literature review

In addition, we employed convenience sampling to identify insights from machine learning, statistics, psychology, economics, biology and other relevant fields. Drawing from academic papers, industry reports and established frameworks, we selected literature based on relevance to general-purpose AI evaluation challenges, prioritising sources addressing internal validity, external validity and reproducibility. Rather than conducting an exhaustive systematic review, we leveraged existing expertise and knowledge of seminal works to address current gaps in GPAI evaluation methodology.

## Limitations

Our methodology has several key limitations. First, relying on a literature review approach itself restricts us to emerging practices that have been documented. Second, the focus on academic literature may not adequately capture real-world implementation challenges faced by practitioners, including time and resource constraints. Future research may a) employ qualitative interviews with practitioners to better capture these perspectives and b) provide a prioritisation of suggestions based on cost-benefit analyses to better address optimal resource allocation under constraints. Third, the December 2024 cutoff date limits the inclusion of evaluation approaches developed for most recent AI models and capabilities in this rapidly evolving field. Fourth, the interdisciplinary knowledge literature review may overemphasise practices from fields in which the authors hold more expertise, including biology, economics, machine learning, psychology and statistics.

## Appendix B: Tagging criteria

**Internal validity (I):** A suggestion may promote internal validity if it increases or helps define the extent to which evaluation results capture the cause-effect relationship between the subject of interest and the outcome metric. This includes suggestions that define the subject of interest and research question, control for confounding variables or improve measurement precision.

**External validity (E):** A suggestion may promote external validity if it increases or helps define the extent to which evaluation results generalise beyond the immediate evaluation environment. This includes suggestions that define or enhance population representativeness, clarify real-world applicability or improve robustness across varied deployment scenarios.

**Reproducibility (R):** A suggestion may promote reproducibility if it increases the likelihood that other researchers could independently obtain consistent results using the same input data, computational methods, code and evaluation conditions. This includes suggestions that enhance methodological transparency, standardise procedures or specify the technical conditions necessary for replication.

## Acknowledgements

The authors thank Michael Aird, Jeff Alstott, Alex Anwyl-Irvine, Kyle Brady, Barbara del Castello, Sunishchal Dev, Casey Dugan, Sarah Gebauer, Ella Guest, Shyam Krishna, Jeffrey Lee, Lorenzo Pacchiardi, Gabriel Sander, Lisa Soder, Brittany Thomas, Christian van Stolk and Sana Zakaria for their support.

## About the authors

**Patricia Paskov** is a Technology and Security Policy fellow at RAND; for more information on the fellowship programme, visit www.rand.org/tasp-fellows. She conducts research on the science of evaluations, international interoperability and societal impacts. She holds a M.Res. and M.S. in Economics.

**Michael Byun** is a Technology and Security Policy fellow at RAND. He conducts technical research on AI evaluations and other AI policy topics. He holds a B.S. in Computer Science.

**Kevin Wei** is a Technology and Security Policy fellow at RAND. Wei conducts research on the science of AI evaluations and on the governance of advanced AI systems. They have an M.S. in Machine Learning and an M.A. in Global Affairs.

**Toby Webster** is a senior research analyst at RAND Europe in the Science and Emerging Technologies Team. He conducts policy research on the intersection of AI and synthetic biology and on biosurveillance. He is also a practicing emergency physician with an MBBS and a BA in Experimental Psychology.

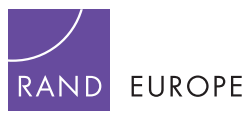

www.randeurope.org